\documentclass[twocolumn,showpacs,superscriptaddress,preprintnumbers,prd]{revtex4}
\usepackage{graphicx,amsmath,dcolumn,bm}
\usepackage{feynmp}
\usepackage{epic,eepic}
\usepackage[dvipdfm]{pict2e}

\def\lsim{\mathrel{\rlap{\lower4pt\hbox{\hskip1pt$\sim$}}
    \raise1pt\hbox{$<$}}}

\usepackage{color}

\definecolor{mycolor}{rgb}{0.6,0.0,0.4}

\begin{document}
\preprint{KEK-TH-1589, J-PARC-TH-0021}
\title{Tomography of exotic hadrons in high-energy exclusive processes}
\author{H. Kawamura}
\affiliation{KEK Theory Center,
             Institute of Particle and Nuclear Studies \\
             High Energy Accelerator Research Organization (KEK) \\
             1-1, Ooho, Tsukuba, Ibaraki, 305-0801, Japan}
\author{S. Kumano}
\affiliation{KEK Theory Center,
             Institute of Particle and Nuclear Studies \\
             High Energy Accelerator Research Organization (KEK) \\
             1-1, Ooho, Tsukuba, Ibaraki, 305-0801, Japan}
\affiliation{
             J-PARC Branch, KEK Theory Center,
             Institute of Particle and Nuclear Studies, KEK \\
           and
           Theory Group, Particle and Nuclear Physics Division, 
           J-PARC Center \\
           203-1, Shirakata, Tokai, Ibaraki, 319-1106, Japan}
\date{January 23, 2014}
\begin{abstract}
We investigated the possibility of determining internal structure of 
exotic hadrons by using high-energy reaction processes, where
quarks and gluons are appropriate degrees of freedom.
In particular, it should be valuable to investigate 
the high-energy exclusive processes which include
generalized parton distributions (GPDs) and
generalized distribution amplitudes (GDAs).
The GPDs and GDAs contain momentum distributions of partons 
and form factors.
We found that the exotic nature appears in momentum distributions 
of quarks as suggested by the constituent-counting rule and 
in the form factors associated with exotic hadron sizes 
and the number of constituents.
Transition GPDs to exotic hadrons could probe such exotic signatures.
We also propose that these exotic signatures should be found
in exclusive production processes of exotic hadrons from $\gamma^*\gamma$ 
in electron-positron annihilation.
For example, the GDAs contain information on a time-like form factor
of the energy-momentum tensor of a hadron $h$. 
We show that the cross section of $e \gamma \to e+h\bar h$
is sensitive to the exotic signature by looking at the $h\bar h$ 
invariant-mass dependence by taking light hadrons, $h=f_0 (980)$ and $a_0 (980)$.
From such GDA measurements, the tomography of exotic hadrons 
becomes possible, for example, by Belle and BaBar experiments and 
by future linear collider.
\end{abstract}
\pacs{13.66.Bc,12.39.Mk}
\maketitle

\section{Introduction}\label{intro}

Almost a half century has passed since the original quark model was
proposed by Gell-Mann and Zweig in 1964. The model suggested that
hadrons are classified into two categories, baryons with three
quarks and mesons with a quark and an antiquark. 
Although there were speculations on other hadron configurations,
such as tetra-quark hadrons of the $qq\bar q\bar q$ configuration 
and penta-quark hadrons of  $qqqq\bar q$, which are not forbidden
by the fundamental theory of strong interactions, 
quantum chromodynamics (QCD), it is rather surprizing that
such exotic hadrons have not been found undoubtedly
for a long time \cite{exotics}.

We are, however, fortunate to start observing some precursory signatures
of exotic hadrons, for example Z(4430), after so many years of theoretical 
and experimental investigations \cite{belle}. 
There are reports on new findings on heavy hadrons with charm 
and bottom quarks \cite{pdg}. However,
even in the low-mass region of 1 GeV, there are controversial hadrons
$f_0 (980)$ and $a_0 (980)$ which could be tetra-quark 
(or $K\bar K$ molecule) hadrons
\cite{spectroscopy-summary,f0-exotic,f0-a0-sk},
and $\Lambda \, (1405)$ which could be 
a penta-quark (or $\bar KN$ molecule) hadron \cite{sk-2013}.
However, internal structure of exotic hadrons is not easily
determined by low-energy measurements on global properties such as
masses and decay widths. Since similar masses and decay widths could be
obtained in ordinary hadron pictures, it is rather difficult to 
entangle various possibilities on internal configurations.

We propose that internal configurations of exotic-hadron candidates
can be determined by high-energy reactions, especially
exclusive processes.
It is particularly noteworthy that quark and gluon degrees of
freedom appear in the high-energy reactions.
Therefore, it is appropriate to use high-energy exclusive 
processes for determining the internal structure of exotic hadrons.
There was a proposal to use fragmentation functions 
for finding the exotic nature by noting the characteristic
differences between favored  and disfavored fragmentation functions 
depending on an internal quark configuration \cite{hkos08}.
For example, the fragmentation functions of an exotic-hadron
candidate $f_0$(980) could be measured by 
the KEK-B factory \cite{seidl}.
We also proposed that the constituent-counting rule in
hard exclusive processes should be useful for determining 
the number of active constituents in exotic hadrons 
by taking $\Lambda (1405)$ production as an example \cite{kks-2013}.
There are also studies on exotic hadrons by  
hard electro-production processes \cite{other-high-energy}.
Such theoretical and experimental studies have not been 
extensively explored yet at high energies although it could 
be considered to be a promising future direction of 
exotic-hadron studies.

On the other hand, internal structure of the nucleon becomes
a new area recently in investigating transverse configuration 
in addition to the longitudinal partonic structure. 
It is the field of ``hadron tomography" to find
the three-dimensional (3D) structure of hadrons. 
It partly originates from the nucleon-spin puzzle in the sense 
that transverse structure, namely orbital angular momenta
of quarks and gluons, needs to be understood because
quark and gluon spin contributions are rather small
for constituting the spin 1/2 of the nucleon.
These 3D picture of the nucleon has been recently investigated
by transverse-momentum-dependent parton distributions (TMDs)
and generalized parton distributions (GPDs) 
\cite{gpd-summary,gpd-gda-summary,transition-gpd,pire-gpd,kss-gpd}.
The TMDs, especially the unpolarized distributions, are sometimes 
called unintegrated parton distribution functions (uPDFs).

These distribution functions contain information on internal
configuration of the nucleon. The longitudinal part contains
the information of longitudinal momentum distributions of partons,
and the transverse part reflects the transverse size.
In this article, we show that these two ingredients could have 
information on exotic hadron structure if the GPDs are
investigated for exotic-hadron candidates.
We also suggest that generalized distribution amplitudes (GDAs),
which are investigated in the $s$-$t$ crossed process to the GPD one,
should be appropriate quantities for determining the internal 
structure of exotic hadrons. Here, $s$ and $t$ are Mandelstam
variables. The GDAs can be investigated in $\gamma^* \gamma$ collisions 
and there are available $e^+e^-$ colliders in the world including
the high-luminosity KEK-B factory for such GDA studies.
The two-photon physics has been studied for investigating
hadron properties \cite{two-photon}, and it can be now used
for the GDA studies.

This article is organized in the following way.
In Sec. \ref{description}, definitions of the GPDs and GDAs are
explained by starting from the basic function of
the Wigner distribution for a hadron.
Then, possible longitudinal momentum distributions and 
form factors are explained for exotic hadrons.
Then, the formalism is shown for calculating exclusive cross sections
for $e + \gamma \rightarrow e + h + \bar h$ to investigate
internal structure of exotic hadron candidates by the GDAs.
Numerical results are shown in Sec. \ref{results}, and they are
summarized in Sec. \ref{summary}.

\section{Formalism}\label{description}

We explain the generalized parton distributions (GPDs) and
generalized distribution amplitudes (GDAs) of hadrons
for investigating internal structure of exotic hadron candidates.
As an example of actual high-energy reaction processes,
an electron-positron annihilation is explained
for investigating the exotic GDAs by exclusive
production processes from $\gamma^* \gamma$.

Although the GPDs and GDAs are well-known quantities for 
nucleon-structure physicists especially in the nucleon-spin community,
they may not be very familiar in low-energy hadron-physics communities.
Therefore, we provide the following introductory explanations on 
the Wigner distributions, GPDs, and GDAs 
in Secs. \ref{wigner}, \ref{gpds}, and \ref{gdas}, respectively.
Our work is to connect the GPDs and GDAs with the exotic-hadron
studies by the forms of momentum distributions of partons, 
form factors, {\it etc.}, and they are explained
in Secs. \ref{exotic-parton}, \ref{transverse-form},
and \ref{results-gda}.

\subsection{Wigner distributions}\label{wigner}

Nucleon structure has been investigated mainly by the form of
deep-inelastic structure functions and elastic form factors.
The structure functions indicate longitudinal momentum 
distributions of partons. In high-energy hadron reactions,
only longitudinal degrees of freedom are usually focused 
by integrating the transverse components. 
However, transverse degrees of freedom are becoming increasingly
important nowadays, for example, in clarifying the origin of
the nucleon spin by orbital angular momenta of partons
and in describing hadron-production processes at hadron
colliders such as RHIC (Relativistic Heavy Ion Collider),
Tevatron, and LHC (Large Hadron Collider) by including
the transverse-momentum dependent parton distributions.
Therefore, the GPDs, TMDs, and uPDFs are now under extensive
and systematic investigations as one of major projects
in hadron physics \cite{gpd-summary,gpd-gda-summary}.
These quantities originate from the Wigner distribution
for the nucleon. In one-dimensional quantum mechanics,
the Wigner distribution is defined by
\begin{align}
W (x,p) = \int d\xi \, e^{i p \, \xi /\hbar} \, 
          \psi^* (x-\xi/2) \, \psi (x+\xi/2) .
\label{eqn:w-1dim-definition}
\end{align}
We consider the simple case of one-dimensional harmonic oscillator
given by the Hamiltonian
$
H = p^2 / (2m) +  m \, \omega^2 x^2 /2 .
$
Solving the Schr\"odinger equation, we obtain the Wigner distribution
\begin{align}
W_n (x,p)  = \frac{(-1)^n}{\pi \hbar} \, e^{-2H/(\hbar \omega)}
            \, L_n \left ( \frac{4 H}{\hbar \omega} \right ) , 
\label{eqn:w-1dim-solution}
\end{align}
where the functions $L_n (x)$ are the Laguerre polynomials.
If the classical limit is taken by $\hbar \rightarrow 0$ and 
$n\rightarrow \infty$, the Wigner function becomes
\begin{align}
W_n (x,p) & \rightarrow \delta \left ( H(x,p) - E_n \right ) 
\ \ \text{as} \, \ \hbar \rightarrow 0, \  n\rightarrow \infty ,
\nonumber\\
E_n & = \hbar \omega \left ( n + \frac{1}{2} \right ), \ \ \ 
n = 0, \, 1, \, 2, \cdot \cdot \cdot .
\label{eqn:w-1dim-classical}
\end{align}
Then, the function $W_n (x,p)$ corresponds to the classical trajectory
in the phase space given by the coordinates $x$ and $p$.
Therefore, the delocalization of the Wigner distribution
indicates quantum effects due to the uncertainty principle.
The Wigner distribution provides information on
quantum states by using the phase-space concept.

In the nucleon case, the Wigner distribution is defined 
for quarks by
\cite{gpd-summary}
\begin{align}
W_\Gamma (x,\vec k_\perp ,\vec r) 
  & = \int \frac{d^3 q}{(2 \pi)^3} \, \left < \, \vec q /2 \, \right |
   \hat w_\Gamma (\vec r, k^+, \vec k_\perp ) 
   \left | \, -\vec q/2 \, \right > ,
\nonumber \\
\hat w_\Gamma (\vec r, k^+, \vec k_\perp ) 
  & = \frac{1}{4\pi} \int d\xi^- \, d^2 \vec\xi_\perp \, 
     e^{i(\xi^- k^+ - \vec\xi_\perp \cdot \vec k_\perp)}
\nonumber \\
  & \ \ \ \ \ \ 
  \left.
  \times \, \bar \psi (\vec r - \xi/2 ) \, \Gamma \, \psi (\vec r + \xi/2 ) 
  \right |_{\xi^+ =0}.
\label{eqn:W-gamma-nucleon}
\end{align}
Here, the $x$ is the Bjorken scaling variable and it should not
be confused with the space coordinate in Eqs. (\ref{eqn:w-1dim-definition}), 
(\ref{eqn:w-1dim-solution}), and (\ref{eqn:w-1dim-classical}),
$\psi$ is the quark field, and
$\Gamma$ is a $\gamma$ matrix which depends on the quark-distribution
type. A gauge link needs to be introduced 
in Eq. (\ref{eqn:W-gamma-nucleon}) for satisfying
the gauge invariance.
The function $W_\Gamma (x,\vec k_\perp ,\vec r)$ provides a complete
quantum mechanical description of the internal nucleon structure
by supplying the phase space distribution at 
the longitudinal momentum fraction $x$, the transverse momentum $\vec k_\perp$,
and the space position $\vec r$.

If this distribution is determined, it means a complete understanding
on the quantum nature of the nucleon substructure from low to high energies. 
However, it is not realistic to determine the Wigner distribution experimentally
because it contains six variables. 
Therefore, the distribution is integrated over some variables
for practical measurements.
Depending on the integral variables, we have following
physics quantities:
\begin{alignat}{2}
& \int d^3 r \, W_\Gamma (x,\vec k_\perp ,\vec r)                      \ \ &
    & \rightarrow \text{TMDs (uPDFs)} ,
\nonumber \\
& \int d^2 k_\perp \, W_\Gamma (x,\vec k_\perp ,\vec r)        \ \ &
    & \rightarrow \text{GPDs} ,
\nonumber \\
& 
\int dx \, d^2 k_\perp \, W_\Gamma (x,\vec k_\perp ,\vec r)  \ \  & 
    & \rightarrow \text{Form factors} ,
\nonumber \\
& \int d^3 r \, d^2 k_\perp \, W_\Gamma (x,\vec k_\perp ,\vec r)       \ \ &
    & \rightarrow \text{PDFs} .
\end{alignat}
As obvious from these expressions,
the TMDs and GPDs provide three-dimensional understanding
of the nucleon by effectively including the PDFs
and transverse distributions.
Now, we introduce basic properties of the GPDs and also 
the GDAs which are the quantities obtained by the $s$ and $t$
channel crossing.

\subsection{Generalized parton distributions}\label{gpds}

\begin{figure}[b!]
\includegraphics[width=0.38\textwidth]{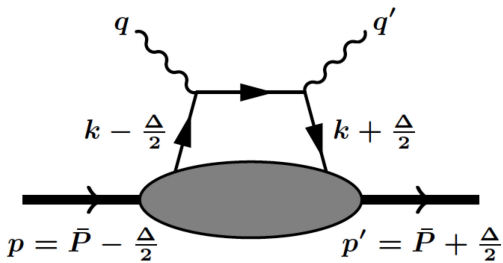}
\vspace{-0.0cm}
\caption{Kinematics for GPDs.}
\label{fig:gpd-1}
\end{figure}

The GPDs have been investigated mainly for the nucleon
in order to clarify the origin of the nucleon spin and
the tomography, namely the determination of
a three-dimensional view of the nucleon substructure.
The GPDs can be measured in various reactions, 
virtual-Compton and meson-production processes in
lepton scattering \cite{gpd-gda-summary,transition-gpd} and 
also in lepton-pair \cite{pire-gpd} and meson production processes 
\cite{kss-gpd} at hadron beam facilities.

The best process to probe the GPDs is the deeply virtual Compton
scattering (DVCS) shown in Fig. \ref{fig:gpd-1}. First, we define
kinematical variables used for describing the DVCS and defining
the GPDs \cite{gpd-gda-summary,freund-nlo-dvcs-2001}.
As shown in Fig. \ref{fig:gpd-1}, $p$ ($q$) and $p'$ ($q'$) are initial 
and final nucleon (photon) momenta, respectively. The initial photon
is a virtual one, whereas the final photon is a real one.
The average nucleon and photon momenta ($\bar P$ and $\bar q$)
and the momentum transfer $\Delta$ are defined by \cite{p-notation}
\begin{align}
\bar P = \frac{p+p'}{2} , \ \ 
\bar q = \frac{q+q'}{2} , \ \ 
\Delta = p'-p = q-q' .
\end{align}
Then, the momentum-transfer-squared quantities are given by
\begin{align}
Q^2 = -q^2 , \ \ 
\bar Q^2 = - \bar q^2 , \ \ 
t = \Delta^2 .
\end{align}
From these quantities, we define 
the generalized scaling variable $x$ and a skewdness parameter $\xi$ by
\begin{align}
x   = \frac{Q^2}{2p \cdot q} , \ \ \ \ 
\xi = \frac{\bar Q^2}{2 \bar P \cdot \bar q} .
\end{align}
The variable $x$ indicates the lightcone momentum fraction 
carried by a quark in the nucleon.
The skewdness parameter $\xi$ or the momentum $\Delta$ indicates
the momentum transfer from the initial nucleon to the final one
or the one between the quarks. 
Using these relations, we can express $\xi$ in terms of $x$, $t$, and $Q^2$ as
\begin{align}
\xi = \frac{x \, [ 1 + t / (2 Q^2) ]}{2 - x \, ( 1 - t / Q^2 ) } 
    \simeq \frac{x}{2 - x} \ \  \text{for} \ Q^2 \gg |t| .
\end{align}

Next, we express these quantities by lightcone variables.
The lightcone coordinates $a^\pm$ are defined by
$a^\pm = (a^0 \pm a^3)/\sqrt{2}$, and
$\vec a_\perp$ is the transverse vector. 
Using these notations, we denote $a=(a^+, \, a^-, \, \vec a_\perp)$.
Neglecting hadron masses, we express the hadron and photon momenta as
\begin{alignat}{2}
& \! \! 
p \simeq \left ( p^+, \, 0, \, \vec 0_\perp \right ) , \ \ \, &
& p' \simeq \left ( {p'}^+, \, 0, \, \vec 0_\perp \right ) ,
\nonumber \\
& \! \! 
q \simeq \left ( -x p^+, \, \frac{Q^2}{2xp^+}, \, \vec 0_\perp \right ) , \ \ \, &
& q' \simeq \left ( 0,     \,  \frac{Q^2}{2xp^+}, \, \vec 0_\perp \right ) .
\end{alignat}
Here, the approximate equations are obtained due to the relation
$(p^+)^2 , \, Q^2 \gg M^2, \, |t|$.
The momentum conservation indicates ${p'}^+ \simeq (1-x) p^+$.
Then, the skewdness parameter $\xi$ is expressed by the lightcone momenta as
\begin{align}
- \frac{\Delta^+}{2\bar P^+}
= \frac{p^+ - {p'}^+}{p^+ + {p'}^+}
= \frac{x}{2 - x}
\simeq \xi
\ \  \text{for} \ Q^2 \gg |t| .
\end{align}
These variables $x$, $\xi$, and $t$ are used for expressing the GPDs.

The GPDs for the nucleon are given by off-forward matrix elements
of quark and gluon operators with a lightcone separation 
between nucleonic states \cite{gpd-summary,gpd-gda-summary}.
The quark GPDs are defined by
\begin{align}
 & \! \! \! \! \! \! \! \! 
 \int\frac{d y^-}{4\pi}e^{i x \bar P^+ y^-}
 \left< p' \left| 
 \overline{\psi}(-y/2) \gamma^+ \psi(y/2) 
 \right| p \right> \Big |_{y^+ = \vec y_\perp =0}
\nonumber \\
 & \! \! \! \! \! \! \! \! \! \! \! \!
 = \frac{1}{2  \bar P^+} \, \overline{u} (p') 
 \left [ H_q (x,\xi,t) \gamma^+
     + E_q (x,\xi,t)  \frac{i \sigma^{+ \alpha} \Delta_\alpha}{2 \, M}
 \right ] u (p) .
\label{eqn:gpd-n}
\end{align}
Here, $\sigma^{\alpha\beta}$ is defined by 
$\sigma^{\alpha\beta}=(i/2)[\gamma^\alpha, \gamma^\beta]$
and $\psi(y/2)$ is the quark field. The functions $H_q (x,\xi,t)$ and 
$E_q (x,\xi,t)$ are the unpolarized GPDs for the nucleon.
The Dirac spinor for the nucleon is denoted as $u (p)$.
There are also gluon GPDs $H_g (x,\xi,t)$ and $E_g (x,\xi,t)$
for the nucleon \cite{gpd-gda-summary}, but they are not 
used in this article for investigating internal structure
of exotic-hadron candidates.

There are three important properties for the GPDs.
First, the nucleonic GPDs $H (x,\xi,t)$ become usual PDFs 
for the nucleon in the forward limit ($\Delta,\, \xi,\, t \rightarrow 0$):
\begin{equation}
           H_q (x, 0, 0) = q(x) ,
\label{eqn:gpd-pdf}
\end{equation}
where $q(x)$ is an unpolarized parton distribution function (PDF)
in the nucleon.
Second, their first moments are the form factors of the nucleon:
\begin{align}
\! \! \! \! \!
\int_{-1}^{1} dx            H_q(x,\xi,t)  = F_1 (t), \ 
\int_{-1}^{1} dx            E_q(x,\xi,t)  = F_2 (t),
\label{eqn:gpd-form}
\end{align}
where $F_1 (t)$ and $F_2 (t)$ are Dirac and Pauli form factors.
Third, a second moment gives a quark orbital-angular-momentum 
contribution ($L_q$) to the nucleon spin:
\begin{align}
   J_q & = \frac{1}{2} \int dx \, x \, [ H_q (x,\xi,t=0) +E_q (x,\xi,t=0) ]
\nonumber \\
       & = \frac{1}{2} \Delta q + L_q ,
\label{eqn:Jq}
\end{align}
where $\Delta q$ is the quark-spin contribution and $J_q$ is
the total angular-momentum of quarks. 
This equation indicates that the nucleonic GPDs are important 
for clarifying the origin of the nucleon spin because
quark and gluon spin contributions seem to be rather small
for explaining the nucleon spin.
From these properties in Eqs. (\ref{eqn:gpd-pdf}) and (\ref{eqn:gpd-form}), 
the GPDs contain both information on the longitudinal momentum distributions
and transverse structure of the nucleon as the form factors.
These facts indicate that the GPDs are valuable quantities
for understanding basic properties of the nucleon from low to high energies.
On the other hand, it is interesting to use these quantities for
exotic-hadron studies because they have information on internal
structure of hadrons.

\subsection{Generalized distribution amplitudes}\label{gdas}

\begin{figure}[t!]
\includegraphics[width=0.35\textwidth]{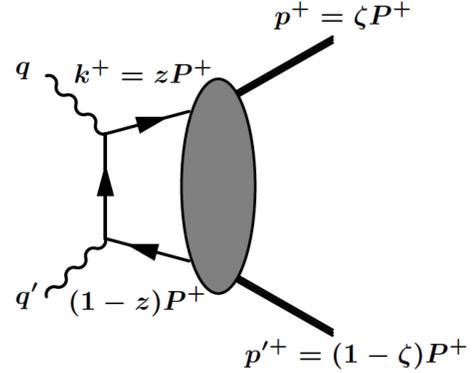}
\vspace{-0.0cm}
\caption{Kinematics for GDAs. The process is a crossed channel
         to the GPD in Fig. \ref{fig:gpd-1}.}
\label{fig:gda-1}
\end{figure}

The GDAs are defined in the same manner with the GPDs 
in the $s$-$t$ crossed channel as shown in Fig. \ref{fig:gda-1}.
They describe the production of a hadron pair $h\bar h$ from a $q\bar q$ 
or gluon pair. 
First, we define kinematical variables for describing the 
$\gamma^* \gamma \rightarrow h \bar h$ process
\cite{gpd-gda-summary,muller-1994,diehl-2000,ee-eerhorho}.
The final hadron momenta are $p$ and $p'$ as shown in Fig. \ref{fig:gda-1},
$P$ is their total momentum: $P=p+p'$ \cite{p-notation}, 
and $k$ is the quark momentum.
The momenta $q$ and $q'$ are initial photon momenta.
The photon with the momentum $q$ is a virtual one with a typical scale
larger than the QCD scale $\Lambda$,
so that the factorization of the process
into a soft part and a hard process should be valid
\cite{factorization}.
Another photon with the momentum $q'$ is considered to be a real one:
\begin{align}
Q^2 = - q^2 \gg \Lambda^2 , \ \ \ 
{q'}^2 = 0 .
\end{align}
The center-of-mass (c.m.) energy squared is $s$ which is equal to
the invariant-mass squared $W^2$ of the final hadron pair:
\begin{align}
s= (q+q')^2 = (p+p')^2 = W^2 .
\end{align}

The variables $x_\gamma$ and $\zeta$ are defined by
\begin{align}
x_\gamma = \frac{Q^2}{2q\cdot q'}
  = \frac{Q^2}{Q^2+W^2}
 , \ \ \ 
\zeta = \frac{p \cdot q'}{P \cdot q'}.
\end{align}
In the center of mass system of two final hadrons, the photon and
hadron momenta are expressed as
\begin{align}
q  & = \left ( q^0,      \, 0, \, 0, \,  |\vec q \,| \right ) , \ \ \ 
q' = \left ( |\vec q \,|, \, 0, \, 0, \, -|\vec q \,| \right ) ,
\nonumber \\
p  & = \left ( p^0,      \,  |\vec p \,| \sin \theta, \, 0, \,  
                             |\vec p \,| \cos \theta \right ) ,
\nonumber \\
p' & = \left ( p^0,      \, -|\vec p \,| \sin \theta, \, 0, \,  
                            -|\vec p \,| \cos \theta \right ) ,
\label{eqn:qqpp}
\end{align}
where $\theta$ is the scattering angle in the c.m. frame.
Then, the variable $\zeta$ becomes 
\begin{align}
\zeta = \frac{p \cdot q'}{P \cdot q'}
      = \frac{p^+}{P^+} = \frac{1+\beta \cos\theta}{2} ,
\label{eqn:zeta-frac}
\end{align}
where $\beta$ is the velocity of a hadron
\begin{align}
\beta = \frac{|\vec p \, |}{p^0} = \sqrt{1-\frac{4 m_h^2}{W^2}} ,
\label{eqn:beta}
\end{align}
Here, $m_h$ is the mass of a final hadron $h$.
Equation (\ref{eqn:zeta-frac}) indicates that 
the variable $\zeta$ is the lightcone momentum ratio
for a hadron in the hadron pair.
Next, as for the lightcone momentum ratio for a quark
in the hadron pair, we use the variable $z$ which is 
given by
\begin{align}
z = \frac{k \cdot q'}{P \cdot q'} = \frac{k^+}{P^+} ,
\end{align}
The GDAs can be expressed in terms of these variables $z$, $\zeta$, and $s$.

The GDAs are defined by the same lightcone operators as the GPDs
between the vacuum and the hadron pair $h\bar h$ instead of 
initial and final hadron states for the GPDs \cite{gpd-gda-summary}:
\begin{align}
& \Phi_q^{h\bar h} (z,\zeta,s) 
= \int \frac{d y^-}{2\pi}\, e^{i (2z-1)\, P^+ y^-}
   \nonumber \\
& \,  
  \times \langle \, h(p) \, \bar h(p') \, | \, 
 \overline{\psi}(-y/2) \gamma^+ \psi(y/2) 
  \, | \, 0 \rangle \Big |_{y^+=\vec y_\perp =0} \, ,
\end{align}
which is a quark GDA. The gluon GDA is defined in the same way
\cite{gpd-gda-summary}.

By considering the kinematical region of $Q^2 \gg W^2, \, \Lambda^2$,
the process $\gamma^* \gamma \rightarrow h \bar h$ is factorized
into two parts as shown in Fig. \ref{fig:gda-1}: 
a hard part described by photon interactions with quarks
and a soft part given by the GDAs.
In the same way, the $\gamma^* h \rightarrow \gamma h'$ is factorized
at $Q^2 \gg -t, \, \Lambda^2$
into the hard part of quark interactions and the soft one given
by the GPDs as shown in Fig. \ref{fig:gpd-1}.
Once the factorization can be applied for both processes, 
the GDAs are related to the GPDs by the $s$-$t$ crossing.
The $\gamma^* \gamma$ process $\gamma^* \gamma \rightarrow h\bar h$
at large $Q^2$ and $W^2 \ll Q^2$ should be related
to the virtual Compton scattering on $h$
($\gamma^* h \rightarrow \gamma h$) at large $Q^2$ and $-t \ll Q^2$
by the $s$-$t$ crossing.
The crossing means to move the final state $\bar h$ ($p'$) 
to the initial $h$ ($p$), which indicates that the momenta ($p$, $p'$)
of the GDAs should be replaced by ($p'$, $-p$) in the GPDs.
It corresponds to the variable changes \cite{gpd-gda-summary,diehl-2000}:
\begin{align}
z \leftrightarrow \frac{1-x/\xi}{2}, \ \ \ \ 
\zeta \leftrightarrow \frac{1-1/\xi}{2}, \ \ \ \ 
W^2 \leftrightarrow t.
\label{eqn:variables-relation}
\end{align}
These relations indicate that the GDAs are related to the GPDs by
the relation:
\begin{align}
& \! \!  
\Phi_q^{h\bar h} (z,\zeta,W^2) 
\nonumber \\
& \! \! 
\longleftrightarrow
H_q^h \left ( x=\frac{1-2z}{1-2\zeta},
            \xi=\frac{1}{1-2\zeta}, t=W^2 \right ) .
\label{eqn:gda-gpd-relation}
\end{align}

Since there are more theoretical and experimental studies in the GPDs,
this correspondence may seem to be convenient to estimate the GDAs 
for exotic hadrons by using an appropriate and simple function 
for the GPDs as explained in Sec. \ref{gpd-exotics}.
However, we notice in Eq. (\ref{eqn:gda-gpd-relation})
that the GDAs correspond to the GPDs in the 
{\it unphysical region} \cite{personal}.
For example, the relation $1-2\zeta =1/\xi$ with $ | 1-2\zeta | \le 1$
indicates $|\xi| \ge 1$, which is an unphysical region of the GPDs.
Furthermore, $x=(1-2z)/(1-2\zeta)$ suggests that $x$ could be larger
than one ($|x| \ge 1$) depending on $z$ and $\zeta$, and 
$t$ is $t=s \ge 0$ although it satisfies $t<0$ in the spacelike reaction. 
These are also unphysical regions. Therefore, it is not straightforward how
the GPDs could be used for the studies of GDAs.
It is also noteworthy that Eq. (\ref{eqn:gda-gpd-relation})
indicates the kinematical region $|\xi| \ge |x|$ which is specifically
called as the Efremov-Radyushkin-Brodsky-Lepage (ERBL) region.
Therefore, the correspondence of the GPDs and GDAs could be
investigated by studying the kinematical region:
\begin{align}
0 \le |x| < \infty, \ \ 
0 \le |\xi| < \infty, \ \ 
|x| \le |\xi| , \ \ 
t \ge 0 ,
\label{eqn:gda-gpd-kinematics}
\end{align}
which is not necessarily the usual physical region of the GPDs.
In any case, it is an interesting topic for future investigations
on the correspondence between the GPDs and GDAs by considering
the $s$-$t$ crossing relation.

\subsection{Cross section of {\boldmath$e + \gamma \rightarrow e + h \bar h$}
            and GDAs}
\label{cross}

\begin{figure}[t!]
\includegraphics[width=0.45\textwidth]{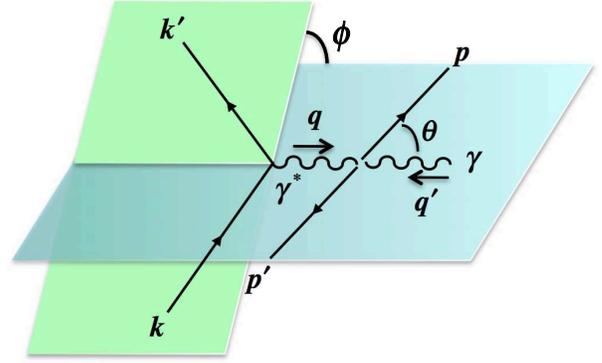}
\vspace{-0.0cm}
\caption{(Color online) Kinematics of 
    the $e + \gamma \rightarrow e + h \bar h$ process.}
\label{fig:cross-kinematics}
\end{figure}

Although the cross section of the two-photon process 
$\gamma^* \gamma \rightarrow h\bar h$ could be used for comparison
with experimental data, we show the cross section
of $e + \gamma \rightarrow e + h \bar h$ in this work.
They are simply related by a kinematical factor, so either
cross section can be shown theoretically for comparing them with
experimental data \cite{uehara}
because we ignore the bremsstrahlung process
as explained after Eq. (\ref{eqn:eps-y}).
The bremsstrahlung process does not exist for neutral-meson
pair productions.

The kinematics of this process is shown in Fig. \ref{fig:cross-kinematics}.
The initial and final electron momenta are denoted as $k$ and $k'$,
the virtual and real photon momenta are $q$ and $q'$, respectively,
and the final hadron momenta are $p$ and $p'$. The scattering angle
for the final hadron in the center of mass of the final hadrons
is $\theta$, and the scattering plane of the electron has the angle
$\phi$ with respect to the reaction plane of
$\gamma^* \gamma \rightarrow h\bar h$ as shown in the figure
\cite{diehl-2000}.
The cross section for $e\gamma  \to e h \bar h$ is given by
\begin{align}
d\sigma  
& = \frac{1}{{4\sqrt {{(k \cdot q')^2} - {k^2}{{q'}^2}} }}
\mathop {\bar \Sigma }
\limits_{{\lambda _\gamma },{\rm{ }}{\lambda _e},{\rm{ }}{{\lambda '}_e}}
{\left| {M(e\gamma  \to e' h \bar h )} \right|^2}d{\Phi _3} ,
\nonumber \\
& d{\Phi _3} = \frac{{{d^3}p}}{{2{E_h }{{(2\pi )}^3}}}
\frac{{{d^3}p'}}{{2{E'_{\bar h} }{{(2\pi )}^3}}}
\frac{{{d^3}k'}}{{2{E'_e}{{(2\pi )}^3}}}
\nonumber \\
& \ \ \ \ \ \ \ \ \ \ \ \ \ \times
{(2\pi )^4}{\delta ^4}(k + q' - p - p' - k') ,
\label{eqn:cross-1}
\end{align}
where $d{\Phi _3}$ is the three-body phase space factor,
and the summation indicates the spin averages.
The center-of-mass-energy square is given by
\begin{align}
s_{e\gamma} = (k+q')^2 = 2 k \cdot q',
\end{align}
by neglecting the electron mass.

The matrix element for $e\gamma  \to e' h \bar h$ is given by
\begin{align}
iM(e\gamma  \to e' h \bar h ) & = 
\bar u(k') \left( { - ie{\gamma _\mu }} \right)u(k)
\nonumber \\
&  \ \  \times
\frac{{ - i{g^{\mu \nu }}}}{{{q^2}}}
\left( { - i{T_{\nu \rho }}} \right){\varepsilon ^\rho }(q') ,
\end{align}
where ${\varepsilon ^\rho }(q')$ is the polarization vector
of the real photon. The polarization vectors of the photons
are given by
\begin{align}
& \varepsilon _\mu ^{( \pm )}(q) 
= \frac{1}{{\sqrt 2 }}\left( {0,
     {\rm{ }} \mp 1,{\rm{ }} - i,{\rm{ }}0} \right),{\rm{  }}
\varepsilon _\mu ^{(0)}(q) 
= \frac{1}{{\sqrt 2 }}\left( {\left| {\vec q} \, \right|,
     {\rm{ }}0,{\rm{ }}0,{\rm{ }}{q^0}} \right) ,
\nonumber \\
& \varepsilon _\mu ^{( \pm )}(q') 
= \frac{1}{{\sqrt 2 }}
\left( {0,{\rm{ }} \mp 1,{\rm{ }} + i,{\rm{ }}0} \right) .
\end{align}
The hadron tensor for $\gamma^* \gamma \to h \bar h$ is defined by
\begin{align}
\! \! \! \! \!
{T_{\mu \nu }} = i\int {{d^4}\xi } {\rm{ }}{e^{ - i\xi  \cdot q}}
\left\langle {h (p) \, \bar h (p') \left| 
{TJ_\mu ^{em}(\xi )J_\nu ^{em}(0)} \right|0} \right\rangle .
\end{align}
Considering a reaction process with the condition $Q^2 \gg \Lambda^2$,
we factorize it into a hard part with the photon interactions
to produce a $q\bar q$ pair and a soft part for producing 
a $h\bar h$ pair from the $q\bar q$ pair as shown in 
Fig. \ref{fig:gda-1}. Namely, it is expressed by the summation
of the quark GDAs $\Phi _q^{h \bar h}$ 
in the leading order of $\alpha_s$ by neglecting gluon GDAs as
\cite{diehl-2000}
\begin{align}
\! \! \! \!
& 
T_{\mu \nu } =  - g_{T\mu \nu }^{}{e^2} 
\sum\limits_q {\frac{{e_q^2}}{2}} 
\int_0^1 {dz} \frac{{2z - 1}}{{z(1 - z)}}
\Phi _q^{h \bar h}(z,\zeta ,{W^2}) ,
\nonumber \\
& \Phi _q^{h \bar h}(z,\zeta ,{W^2})
 = \int {\frac{{d{x^ - }}}{{2\pi }}} 
 {e^{ - iz{P^ + }{x^ - }}} 
\nonumber \\
& \ \ \ \ \ \ \ \times
{\left\langle {h (p) \, \bar h (p') \left| 
{\bar q({x^ - }){\gamma ^ + }q(0)} \right|0} \right\rangle _{{x^ + } 
= 0,{{\vec x}_ \bot } = 0}} ,
\label{eqn:hadron-tensor-LT}
\end{align}
where $g_{T\mu \nu }$ is defined by
\begin{alignat}{2}
g_{T\mu \nu } & = -1  & \ \  & \text{for $\mu=\nu=1, \ 2$} ,
\nonumber \\
              & = 0  & \ \     & \text{for $\mu$, $\nu=$others} .
\end{alignat}
Equation (\ref{eqn:hadron-tensor-LT}) is the leading-twist
expression which is valid at $Q^2 \gg \Lambda^2$, and
higher-twist and higher-order terms are neglected.

We define helicity amplitudes by
\begin{align}
{A_{i  j}} & = \frac{1}{e^2} \,
\varepsilon _\mu ^{( i )}(q) \, \varepsilon _\nu ^{( j )}(q') \,
{T^{\mu \nu }} , 
\nonumber \\
& \ \ \ \ \ \ \ 
i=-,\ 0, \ +, \ \ 
j=-,\ + .
\end{align}
Using the relation 
$\varepsilon _\mu ^{( + )}(q)\varepsilon _\nu ^{( + )}(q')g_T^{\mu \nu } = - 1$,
we obtain, for example, $A_{++}$ as
\begin{align}
A_{++} = \sum\limits_q {\frac{{e_q^2}}{2}} \int_0^1 {dz} 
\frac{{2z - 1}}{{z(1 - z)}}\Phi _q^{h \bar h}(z,\zeta ,{W^2}) ,
\label{eqn:A++}
\end{align}
and another one $A_{- -}$ is also the same, $A_{- -}=A_{+ +}$.
In this way, the matrix element part of the cross section becomes
\begin{align}
\mathop {\bar \Sigma }\limits_{{\lambda _\gamma },{\rm{ }}
{\lambda _e},{\rm{ }}{{\lambda '}_e}} {\left| M \right|^2} 
= \frac{{64{\pi ^3}\alpha _{}^3}}{{{Q^2}(1 - {\rm{ }}
\varepsilon )}}{\left| {{A_{+ +}}} \right|^2} .
\label{eqn:matrix-1}
\end{align}
Here, the $\varepsilon$ is the ratio of 
longitudinal and transverse fluxes of the virtual photon,
and a virtual-photon cross section is generally written as
$\sigma \propto (\sigma_T + \varepsilon \sigma_L )$
by the transverse and longitudinal cross sections $\sigma_T$
and $\sigma_L$, respectively \cite{trans-long}.
The $\varepsilon$ is expressed by another variable $y$ as
\begin{align}
\varepsilon = \frac{1-y}{1-y+y^2/2} , \ \ 
y \equiv \frac{q \cdot q'}{k \cdot q'}
= \frac{Q^2+W^2}{s_{e\gamma}} .
\label{eqn:eps-y}
\end{align}
In principle, other helicity amplitudes such as $A_{- +}$ and $A_{0 +}$
should contribute to the matrix element in Eq. (\ref{eqn:matrix-1}).
However, the leading-twist and leading-order expression is introduced in 
Eq. (\ref{eqn:hadron-tensor-LT}), so that it is expressed
consequently only by the amplitude $A_{+ +}$.
As explained in Ref. \cite{diehl-2000}, relative 
contributions of $\gamma^* \gamma$ and bremsstrahlung 
processes are given by $1/(Q^2(1-\varepsilon))$ and 
$2 \beta^2/(W^2 \varepsilon)$, respectively. 
The bremsstrahlung part is small in our kinematics 
of Sec. \ref{results}, so that it is neglected.
Next, we calculate the phase space factor in Eq. (\ref{eqn:cross-1})
by using the polar and azimuthal angles $\theta$ and $\phi$
in the the c.m. frame of $h \bar h$ as defined in Eq. (\ref{eqn:qqpp})
and Fig. \ref{fig:cross-kinematics}.
It is obtained as
\begin{align}
d{\Phi _3} 
= \frac{\beta }{{8 \cdot 64{\pi ^4}s_{e\gamma}}}d{Q^2}d{W^2}d(\cos \theta) d\phi .
\label{eqn:3-phase}
\end{align}
Substituting Eqs. (\ref{eqn:matrix-1}) and (\ref{eqn:3-phase})
into Eq. (\ref{eqn:cross-1}), we finally obtain the cross section
\begin{align}
\frac{{d\sigma }}{{d{Q^2}d{W^2}d\cos \theta}} 
= \frac{\beta \, \alpha ^3}{8 s_{e\gamma}^2 \, Q^2 (1 - \varepsilon ) }
{\left| {{A_{+ +}}(\zeta ,{W^2})} \right|^2} ,
\label{eqn:corss-2}
\end{align}
where the cross section is integrated over $\phi$.
The variable $\zeta$ and the angle $\theta$ are related by
Eq. (\ref{eqn:zeta-frac}),
and $\beta$ is given in Eq. (\ref{eqn:beta}).
We should note that the cross section generally depends on
the angle $\phi$ \cite{diehl-2000}.
However, the higher-twist terms in the cross sections are
dropped in Eq. (\ref{eqn:corss-2}), which originally comes
from the leading-twist expression of 
Eq. (\ref{eqn:hadron-tensor-LT}).
It leads to the cross section, which is independent of 
the azimuthal angle $\phi$.
The purpose of this work is to investigate basic signatures
of exotic hadrons in the two-photon processes by neglecting
small perturbative QCD corrections. In future, the higher-twist
corrections could become necessary for the detailed comparison
between theory and experimental data.

\section{Results}\label{results}
\label{results}

\subsection{Generalized parton distributions  \\ of exotic hadrons}
\label{gpd-exotics}

For numerical analysis of cross sections with GPDs and GDAs,
realistic functions are needed for these distributions.
A full theoretical information is not yet available for these 
distributions. However, there are some theoretical studies 
on the appropriate parametrization for the GPDs although 
they are scarce for the GDAs.

Because the longitudinal part of the GPDs indicates 
the longitudinal momentum distributions of partons
and the transverse part is related to the two-dimensional
form factors, the simplest idea is to assume a GPD
as the corresponding PDF multiplied by a form factor.
However, it is too simple to describe the realistic hadron.
The large-$x$ distribution is related to a hard valence-quark
with large momentum and it should be confined in a small space region,
namely the transverse distribution is also confined in 
a small region. On the other hand, small-$x$ partons are rather loosely 
distributed in a larger space. Namely, the two-dimensional form factors
are different in small- and large-$x$ regions.
The simple multiplication of the form factor and the PDF
cannot reflect this nature.
In Ref. \cite{gpd-paramet}, the mixed functional form 
is proposed for the nucleon by assuming a Gaussian 
distribution for the transverse part to improve such an issue.

We may use this kind of idea for exotic hadron studies
by supplying the longitudinal momentum distribution $f_n (x)$ of
an exotic hadron with $n$ valence quarks
and the transverse form factor $F_n^h (t, x)$:
\begin{align}
H_q^h (x,\xi=0,t)= f_n (x) \, F_n^h (t, x) ,
\label{eqn:gpd-paramet1}
\end{align}
for valence quarks.
However, the ERBL region $|\xi| > x$ cannot be described
by this functional form because the process is related to 
a meson distribution amplitude. 
A more sophisticated function type is used in 
Ref. \cite{radyushkin-paramet} by including the ERBL region.
Since this article is a first step toward GPDs and GDAs of
exotic hadrons, we simply considered the simple case of
Eq. (\ref{eqn:gpd-paramet1}) in the following discussions.
Significant efforts are still needed for realistic tomography
of exotic hadrons by using the GPDs and GDAs.

\subsection{Parton distribution functions of exotic hadrons}
\label{exotic-parton}

The longitudinal momentum distributions of partons have been investigated
extensively by various processes including charged-lepton and neutrino
deep inelastic scattering (DIS), Drell-Yan processes, and W, $Z$, jet, 
and hadron productions. Now, the details of the PDFs are accurately
determined except for extreme kinematical conditions.
The parton distributions in the pion have been determined by Drell-Yan
processes. Since the unstable pion cannot be used as a fixed target,
there is no deep inelastic measurement. The exotic hadron candidates
cannot be used either as a fixed target or a beam for the Drell-Yan.
Therefore, the longitudinal momentum distribution cannot be measured
by these usual methods. 
However, let us consider a {\it gedankenexperiment} by which
the PDFs or the GPDs of the exotic hadrons can be obtained
by assuming a stable target.

We explain our guideline for the PDFs of the exotic hadrons 
if they were measured.
We consider only the valence-quark distributions. 
Including PDFs in exotic hadrons, we denote 
the sum of valence-quark distributions as $f_n (x)$,
where $n$ is the number of valence quarks in a hadron, by taking
a suitable low $Q^2$ scale ($Q^2=1 \sim$ a few GeV$^2$)
for the first estimate.
In future, scaling violation may be considered; however, 
it is not the stage to investigate such details of $Q^2$ dependence
because there is the first work on possible PDFs of exotic hadrons.

\begin{figure}[t!]
\includegraphics[width=0.25\textwidth]{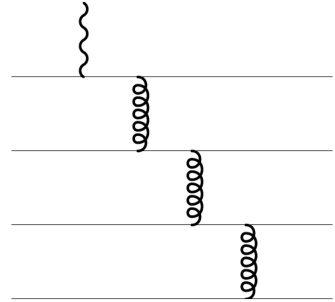}
\vspace{-0.2cm}
\caption{Constituent-counting rule for a tetra-quark hadron.
A three-gluon-exchange process is shown for a typical
exclusive process. There are many other processes depending
how  the gluon lines are attached to the quarks.}
\label{fig:counting-rule}
\end{figure}

If the nucleon consists of three valence quarks and 
if the nucleon momentum is equally shared by the quark momenta,
the distribution has a peak at $x=1/3$: $f_3 (x) \sim \delta (x-1/3)$.
However, the quarks interact with each other by exchanging gluons,
so that their momenta are redistributed.
The distribution should vanish at $x=1$ kinematically, and
the $x$ dependence at $x \rightarrow 1$ is theoretically
described by the constituent-counting rule, obtained
by hard gluon exchanges between quarks for exclusive processes
as shown in Fig. \ref{fig:counting-rule}
for a tetra-quark hadron.
It leads to the $x$ dependence $(1-x)^\beta$ at $x \rightarrow 1$.
Including a polynomial form of $x^\alpha$, 
which is motived, for example, by the Regge theory \cite{trans-long}, 
at smaller $x$, we have a simple functional form 
of the parton distributions as
\begin{align}
f_n (x) = C_n \, x^{\alpha_n} \, (1-x)^{\beta_n} ,
\label{eqn:fn}
\end{align}
for hadrons including exotic ones.

The three constants $C_n$, $\alpha_n$, and $\beta_n$ are
constrained by the valence-quark number $n$
and the momentum $\langle x \rangle_q$ carried by the valence quarks:
\begin{align}
\int_0^1 dx \, f_n (x) = n, \ \ \ 
\int_0^1 dx \, x \, f_n (x) = \langle x \rangle_q .
\end{align}
Using the constant $\beta_n$ and the number of constituents $n$,
we have $C_n$ and $\alpha_n$ as
\begin{align}
\! \! \! 
C_n = \frac{\langle x \rangle_q}{B(\alpha_n+2,\beta_n+1)} ,
\ \ 
\alpha_n = \frac{\langle x \rangle_q (\beta_n+2) -n}
                {n-\langle x \rangle_q}.
\label{eqn:constants}
\end{align}
The large-$x$ limit ($x \rightarrow 1$) corresponds to
the exclusive process shown in Fig. \ref{fig:counting-rule}.
There exists the Drell-Yan-West relation between the Dirac form factor
at large $Q^2$ and the structure function $F_2$ at $x \to 1$
\cite{d-y-west-1970,melnitchouk-2005}.
It indicates $F_2 (x) \sim (1-x)^{2\gamma-1}$ at $x \to 1$
for the form-factor function $F_1 (Q^2) \sim 1/(Q^2)^{\gamma}$
at large $Q^2$. This relation was derived before the advent of QCD;
however, the constituent-counting rule was derived in perturbative
QCD by describing the process by hard gluon exchanges typically
shown in Fig. \ref{fig:counting-rule}
\cite{pqcd-counting}. 
As explained in Ref. \cite{kks-2013}, 
the large momentum provided by the hard photon should be shared
by four quark constituents by exchanging three hard gluons
so that the initial and final states are the same.
From the hard gluon and quark propagators together with 
other kinematical factors, the process indicates that 
the factor $\gamma$ expressed by the number of active constituents $n$
as $\gamma = n-1$ in the form factor at large $Q^2$.
It leads to the overall factor of $(1-x)^{\beta_n}$, 
$\beta_n = 2n -3$ \cite{d-y-west-1970}
with the number of valence quarks $n$. 
Because of this number counting, it is called the constituent-counting rule
for structure functions at $x \rightarrow 1$ and also
for the form factors. However, it was pointed out that there is 
an additional spin factor $\Delta S_z=|S_z^q-S_z^h|$
due to the helicity conservation \cite{pQCD-large-x-pdf}, 
so that the factor becomes $\beta_n = 2n -3+2\Delta S_z$.
It does not matter for the spin-1/2 nucleon because $\Delta S_z=0$; 
however, it affects the pion because of $\Delta S_z=1/2$.
In discussing a spin-1/2 penta-quark hadron candidate
such as $\Lambda (1405)$,
it also does not matter for the unpolarized distributions.
For showing the distribution for 
a tetra-quark hadron, we assume that it is spin-0 particle
so as to have the factor $\Delta S_z=1/2$ by considering,
for example, $f_0 (980)$ and $a_0 (980)$.

\begin{figure}[b!]
\includegraphics[width=0.42\textwidth]{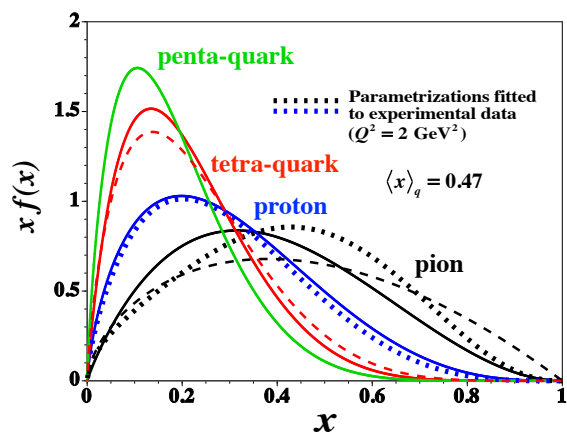}
\vspace{-0.2cm}
\caption{(Color online) PDFs of exotic hadrons in comparison
with parametrizations of pion and proton PDFs.
The solid curves are calculated by Eqs. (\ref{eqn:fn})
and (\ref{eqn:constants}) with $\beta_n = 2n -3+2\Delta S_z$
\cite{pQCD-large-x-pdf}. The spin factor is $\Delta S_z=0$ 
for the proton and penta-quark, whereas it is $\Delta S_z=1/2$
for the pion and tetra-quark. In comparison, the native
estimates with $\beta_n = 2n -3$ \cite{d-y-west-1970} 
are shown by the dashed curves. 
The dotted curves indicate the PDF parametrizations 
at $Q^2$=2 GeV$^2$ for the proton and pion.
}
\label{fig:exotic-pdfs}
\end{figure}

Using Eqs. (\ref{eqn:fn}) and (\ref{eqn:constants}),
we calculate the PDFs for hadrons with the valence-quark
number $n=2,\, 3,\, 4$, or $5$, namely for ordinary 
$q\bar q$-type mesons, $qqq$ type baryons,
tetra-quark and penta-quark hadrons. 
Their valence-quark distributions are shown by the solid 
and dashed curves in Fig. \ref{fig:exotic-pdfs}.
The average momentum fraction $\langle x \rangle_q$ is considered 
to be in the range $0.4 \sim 0.5$, so that
it is taken as $\langle x \rangle_q=0.47$ in the figure.
The distributions have peaks at about $x=1/n$.
For example, there are two valence quarks in the pion, so that 
it has a wide-$x$ distribution with the peak position at $x \sim 1/2$.
As the valence-quark number becomes larger for tetra-quark and
penta-quark hadrons, the hadron momentum is shared by more 
(four or five) valence quarks, so that the distributions becomes
softer as the peak moves to smaller-$x$ locations.

First, we need to confirm that the simple counting-rule
estimates are reasonable in comparison with experimental data.
There exist experimental measurements for the PDFs in the pion
and the proton. The pionic PDFs are determined by the Drell-Yan,
and the proton PDFs are by various measurements including
leptonic DIS. Because the optimum PDFs are determined from
global analyses of many experimental measurements and the PDFs
cannot be directly measured, the experimental data cannot be 
plotted in the figure. Therefore, the typical parametrizations
obtained by the global analyses are shown in 
Fig. \ref{fig:exotic-pdfs}. 

Since the counting-rule PDFs do not contain a specific 
$Q^2$ scale, there is uncertainty how to choose $Q^2$
in such parametrizations. We chose $Q^2$=2 GeV$^2$ 
in Fig. \ref{fig:exotic-pdfs} as a typical $Q^2$ 
which is close to the hadronic scale and yet it is
the region where the perturbative QCD can be applied.
The valence-quark distributions have strong constraints
by the sum rules coming from hadron charges and baryon numbers. 
Therefore, the distribution shapes do not
change drastically depending on $Q^2$ due to 
the sum-rule constraints in addition to the fact that
the PDFs have, in general, weak $Q^2$ dependence 
in the logarithmic $Q^2$ form \cite{Q2-code}.

The PDFs at $Q^2$=2 GeV$^2$ are taken from the next-to-leading order (NLO)
distributions of the fit-3 in Ref. \cite{pion-pdf}
for the pion and in Ref. \cite{proton-pdf} for the proton.
Although the comparison of the results in Fig. \ref{fig:exotic-pdfs}
depends on the choice of $\langle x \rangle_q=0.47$ and the $Q^2$ value,
the counting-rule PDFs are reasonably consistent with the current 
experimental measurements for the pion and proton PDFs,
although the detailed $x$ dependence of
the pion functions with $\beta_\pi =1$ and $2$ is slightly different
from the experimental determination shown by the dotted curve.
Therefore, we believe that the valence-quark distributions
for the tetra-quark and penta-quark hadrons in Fig. \ref{fig:exotic-pdfs}
should be reasonable.

We found that the longitudinal momentum distribution of 
the valence quarks has a characteristic feature that 
the distribution is shifted toward the smaller-$x$ region.
If such a signature is found experimentally, it should be
a key discovery in exotic hadron search because a clear
signature is difficult to be found in low-energy measurements
of hadron masses and decays. However, the exotic hadron
candidates cannot be used for fixed targets and beams,
so that possible methods should be studied by including
transition GPDs \cite{transition-gpd,kss-gpd}.
In this article, we propose that the GPDs and GDAs (and also TMDs)
are appropriate quantities to be investigated because the exotic
hadrons could exists in the final state, rather than the initial
targets and beams, and because the GPDs and GDAs contain
the PDFs in their functions as the longitudinal-momentum
dependence.

\vspace{-0.3cm}
\subsection{Transverse form factors of exotic hadrons}
\label{transverse-form}
\vspace{-0.3cm}

Next, we show typical transverse form factors $F_n^h (t,x)$, 
which is given in Eq. (\ref{eqn:gpd-paramet1}). 
A simple form of this function is given in
as an exponential form \cite{gpd-paramet}:
\begin{align}
F_n^h (t,x) = e^{(1-x) t/(x \Lambda^2)} ,
\end{align}
where $\Lambda$ is cutoff parameter for the transverse momentum.
The transverse density is given by its Fourier transform
\begin{align}
\rho (r_\perp, x) & = \int \frac{d^2 q_\perp}{(2\pi)^2}
e^{-i \vec q_\perp \cdot \vec r_\perp} F_n^h (t=-q_\perp^2,x) 
\nonumber \\
& = \frac{x \, \Lambda^2}{4\pi (1-x)}
e^{-x  \Lambda^2 r_\perp^2 /(4 (1-x))} ,
\label{eqn:rho-2dim}
\end{align}
and the root-mean-square (rms) radius is then expressed 
by the cutoff parameter as
\begin{align}
\left <  r_\perp^2 \right >
= \frac{4(1-x)}{x \Lambda^2} .
\end{align}

As for the pion and nucleon form factors, 
monopole and dipole forms ($1/(|t|+\Lambda^2)$, $1/(|t|+\Lambda^2)^2$)
are usually used, so that the Gaussian form could be too
steep as a function of $q_\perp^2$. The dipole form factor
for the nucleon corresponds to the exponential density distribution,
and the Gaussian form factor does to the Gaussian as shown in
Eq. (\ref{eqn:rho-2dim}). In nuclear physics, such a Gaussian
form factor is realistic on in light nuclei such as $^6$Li
\cite{form-factors}.
In any case, the functional form depends on an exotic hadron
type and it should be the topic of future investigations, 
and we do not step into such details in this work.

\begin{figure}[t!]
\includegraphics[width=0.40\textwidth]{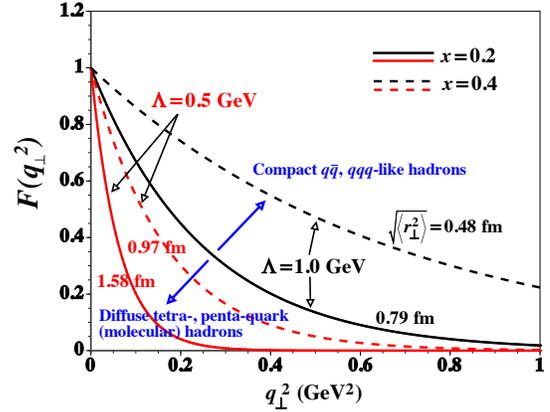}
\vspace{-0.3cm}
\caption{(Color online) Transverse form factors for $x=0.2$, 0.4 and
$\Lambda=0.5$, 1.0 GeV. The r.m.s radius is 
$\sqrt{<r_\perp^2>}=0.48$, 0.79, 0.97, and 1.58 fm for
$(x,\, \Lambda \text{(GeV)})=$(0.4, 1.0), (0.2, 1.0),
(0.4, 0.5), (0.2, 0.5), respectively.}
\label{fig:transverse-form}
\end{figure}

As an example, the transverse form factor is shown 
as a function of $q_\perp^2$ in Fig. \ref{fig:transverse-form}.
The form factors for the cutoff parameters, $\Lambda$=0.5 and 1.0 GeV,
are shown at $x=0.2$ and 0.4.
The rms radii are $\sqrt{<r_\perp^2>}=0.48$, 0.79, 0.97, and 1.58 fm 
for the variable $x$ and the cutoff parameter $\Lambda$,
$(x,\, \Lambda \text{(GeV)})=$(0.4, 1.0), (0.2, 1.0),
(0.4, 0.5), (0.2, 0.5), respectively.
The three-dimensional charge radii are measured as
0.877 fm for the proton and 0.672 fm for $\pi^\pm$ \cite{pdg-kek}.
Because the current transverse radius depends on the longitudinal
momentum fraction $x$, the transverse radius is not equal to
the rms radius. However, the middle region of the four curves
corresponds to the ``ordinary" hadrons consists of $q\bar q$ and
$qqq$ types. On the other hand, the tetra- and penta-quark hadrons,
particularly the molecular type hadrons, have large spacial distributions,
so that the momentum distribution becomes softer as typically indicated
by the curves with $\Lambda$=0.5 GeV.

\vspace{0.2cm}
\noindent
{\bf Summary of exotic signatures in the GPDs}
\vspace{0.1cm}

The simple GPD parametrization of Eq. (\ref{eqn:gpd-paramet1})
indicates that internal configurations of exotic hadrons could
be clarified by finding typical exotic signatures
in the longitudinal and transverse distributions.
\vspace{-0.15cm}
\begin{enumerate}
\item[(1)] Softening of longitudinal momentum distribution: \\ 
As indicated in Fig. \ref{fig:exotic-pdfs}, the longitudinal
momentum distribution, namely the PDF, should shift to smaller
$x$ as it becomes tetra- or penta-quark hadrons as expected
from the quark counting rule and the momentum fraction carried
by quarks.
\vspace{-0.15cm}
\item[(2)] Softening of transverse form factor: \\
The transverse momentum distribution becomes softer for 
tetra- and penta-quark hadrons, especially for molecular-type
hadrons, than the one expected from ordinary compact $q \bar q$ 
or $qqq$ type hadrons.
\end{enumerate}
\vspace{-0.15cm}

\begin{figure}[t!]
\includegraphics[width=0.46\textwidth]{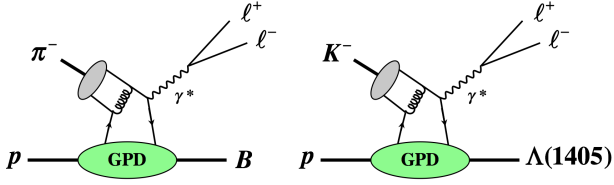}
\vspace{-0.2cm}
\caption{(Color online) Examples of transition GPDs 
by exclusive dimuon production processes with
pion and kaon beams.
$\Lambda (1405)$ is an exotic hadron candidate.}
\label{fig:lepton-pair-gpds}
\end{figure}

However, exotic hadrons are unstable particles and they cannot be used 
as a target in experimental measurements. A possible solution is
to investigate a production process of an exotic hadron, which
is associated with transition GPDs \cite{transition-gpd,kss-gpd} 
from an ordinary hadron such as the proton to an exotic one. 
An exclusive lepton-pair production process 
$\pi + N \to \mu^+ \mu^- +N$ can be used
for investigating the nucleonic GPDs \cite{pire-gpd}.
The exclusive $\mu^+ \mu^-$ production could be valuable also 
for investigating exotic hadrons such as $\Lambda (1405)$
by the form of transition GPDs from the proton to $\Lambda (1405)$
as illustrated in Fig. \ref{fig:lepton-pair-gpds}.
Since high-momentum beams of protons and pions, possibly also
kaons in future, will become available at J-PARC,
an experimental proposal is considered \cite{j-parc-pro-high}.
We expect to investigate such a topic as our future work.
Here, we discuss an alternative method by using the GDAs,
which the exotic hadrons are studied by time-like production
processes.

\vspace{-0.3cm}
\subsection{Internal structure of exotic hadrons by GDAs \\
            in electron-positron annihilation}\label{exotics-e+e-}
\label{results-gda}
\vspace{-0.3cm}

In order to estimate the cross section for $e \gamma \to e' h\bar h$
by Eqs. (\ref{eqn:A++}) and (\ref{eqn:corss-2}),
the actual functions of GDAs need to be introduced. 
There was a work on the GDAs on their impact-parameter dependence
\cite{impact-parameter}.
In our studies, we use a simple function, which satisfies
the sum rules for the quark GDAs for 
the isospin $I=0$ two-meson final states \cite{diehl-2000,gda-sum-rule}:
\begin{align}
\int_0^1 dz \, & \Phi_q^{h\bar h (I=0)} (z,\zeta,W^2) = 0, 
\nonumber \\
\int_0^1 dz \, & (2z -1) \, \Phi_q^{h\bar h (I=0)} (z,\zeta,W^2) 
\nonumber \\
& \ \ 
= - 2 M_{2(q)}^h \zeta (1-\zeta) F_{h(q)} (W^2) ,
\label{eqn:gda-sum-I=0}
\end{align}
where $M_{2(q)}^h$ is the momentum fraction carried by quarks,
and $F_{h(q)} (W^2)$ is a form factor of the quark part
of the energy-momentum tensor
\cite{energy-momentum}.
One may note that the kinematical condition $Q^2 \gg W^2$ is needed
for the factorization of Fig. \ref{fig:gda-1}. If it is satisfied,
the two-meson final state originates from the $q\bar q$ state
from $\gamma^* \gamma$, so that only the $I=0$ state is allowed.
A simple functional form to satisfy these conditions is
\begin{align}
& \Phi_q^{h\bar h (I=0)} (z,\zeta,W^2) 
\nonumber \\
& \ \ \ 
= N_{h(q)} \, z^\alpha (1-z)^\beta (2z-1) \, \zeta (1-\zeta) \, F_{h(q)} (W^2) ,
\label{eqn:gda-paramet}
\end{align}
where the normalization constant is given by
\begin{align}
& \! \! \! 
N_{h(q)} = - \frac{2 M_{2(q)}^h}{B(\alpha+1,\beta+1)}
\nonumber \\
& \! \! \! \! \! \! \! \! \! \!
\times
\frac{(\alpha+\beta+2)(\alpha+\beta+3)}
{(\alpha+1)(\alpha+2)+(\beta+1)(\beta+2)-2(\alpha+1)(\beta+1)} .
\label{eqn:normalization}
\end{align}
In Eq. (\ref{eqn:gda-paramet}), a flavor-independent 
functional form is used except for the normalization factor.

\begin{figure}[t!]
\includegraphics[width=0.38\textwidth]{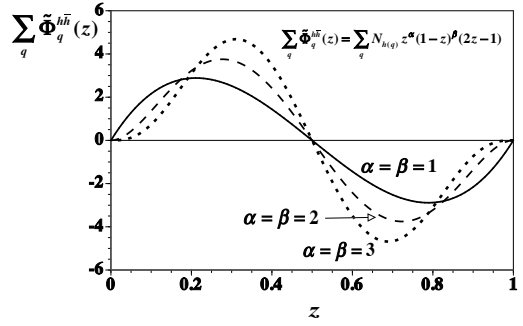}
\vspace{-0.3cm}
\caption{$z$ dependence is shown for the GDA
$\tilde\Phi_q^{h\bar h} (z) = N_{h(q)} z^{\alpha} (1-z)^\beta (2z-1)$
by taking $\alpha=\beta=1, \ 2$, and $3$. They are shown by
the solid, dashed, and dotted curves, respectively.}
\label{fig:phi-z}
\end{figure}

\begin{figure}[t!]
\vspace{-0.3cm}
\includegraphics[width=0.34\textwidth]{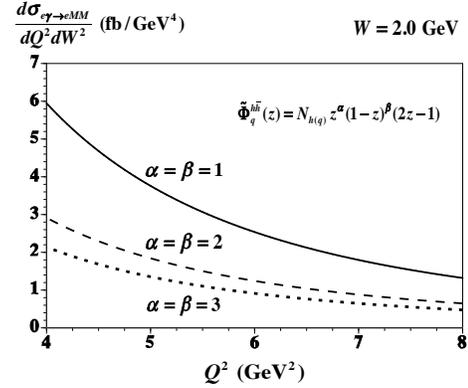}
\vspace{-0.3cm}
\caption{Effect on the cross section $d\sigma/(dQ^2 dW^2)$ 
at $W=$2 GeV from the $z$ dependence in the GDA
$\tilde\Phi_q^{h\bar h} (z) = N_{h(q)} z^{\alpha} (1-z)^\beta (2z-1)$
with $\alpha=\beta=1, \ 2$, and $3$. They are shown by
the solid, dashed, and dotted curves, respectively.
Here, the hadron $h$ is $f_0$ or $a_0$,
and $n=2$ is taken in the form factor of Eq. (\ref{eqn:form-factorw2}).}
\label{fig:cross-phi-z}
\end{figure}

First, as the $x$-dependent PDFs indicated exotic signature
in Fig. \ref{fig:exotic-pdfs}, the $z$-dependent function should
reflect some exotic features because the variables $x$ and $z$ 
are related by Eq. (\ref{eqn:variables-relation}).
However, we could not find a useful relation like the counting
rule at $x \to 1$ in the PDFs for restricting the functional
form. Because small $x$ corresponds to $z \to 1/2$, 
the functional variation for exotic hadrons could correspond
to the variation of the constants $\alpha$ and $\beta$.
Therefore, we took $\alpha=\beta=1$, 2, and 3 to show
its variation. The results are shown for
the $z$ dependent part 
$\sum_q \tilde\Phi_q^{h\bar h (I=0)} (z) 
= \sum_q  N_{h(q)} \, z^\alpha (1-z)^\beta (2z-1)$
in Fig. \ref{fig:phi-z} by taking 
$\sum_q M_{2(q)}^h=0.5$.
Then, their effects on the cross section are shown 
in Fig. \ref{fig:cross-phi-z}.
The form factor of Eq. (\ref{eqn:form-factorw2}) is used
with $n=2$.
As the function $\sum_q \tilde\Phi_q^{h\bar h (I=0)} (z)$ becomes steep
as shown in Fig. \ref{fig:phi-z} with increasing $\alpha$ and $\beta$,
the cross section becomes smaller.
In order to clarify the $z$-dependent function, we need
an estimate of the GDAs based on a theoretical model.

\begin{figure}[t!]
\includegraphics[width=0.35\textwidth]{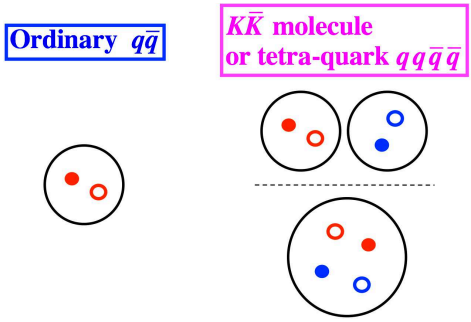}
\vspace{-0.2cm}
\caption{(Color online)
Schematic picture of $f_0$ and $a_0$ mesons, which
are exotic-hadron candidates.
It is likely that they are $K\bar K$ molecule or tetra-quark
hadron. Because their sizes and internal constituents are
different from the ordinary $q\bar q$-type mesons, they
should be distinguished by the hadron tomography by using
the GDAs. }
\label{fig:exotic-picture}
\end{figure}

As for the form factor of the energy-momentum tensor,
we employ a simple form suggested by the constituent-counting rule
in perturbative QCD \cite{BC-1976}:
\begin{align}
F_{h(q)} (W^2) 
= \frac{1}{\left[ 1 + (W^2-4 m_h^2)/\Lambda^2 \right]^{n-1}} ,
\label{eqn:form-factorw2}
\end{align}
where $\Lambda$ is the cutoff parameter, which indicates
the hadron size, $n$ is the number of active constituents,
and it is normalized as $F_{h(q)} (4 m_h^2)=1$.
Here, the phase factor does not exist because the bremsstrahlung
process is neglected.
It is $n=2$ for an ordinary $q\bar q$ hadron, whereas
$n=4$ for a $K\bar K$ molecule or a tetra-quark hadron
as illustrated in Fig. \ref{fig:exotic-picture}.

\begin{figure}[t!]
\includegraphics[width=0.42\textwidth]{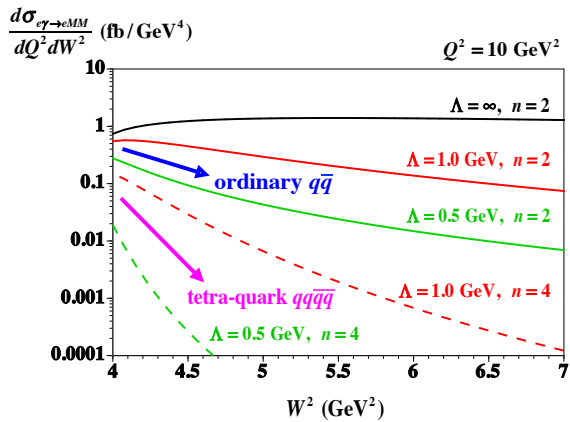}
\vspace{-0.2cm}
\caption{(Color online)
Form-factor effects are shown on the cross section 
$d\sigma/(dQ^2 dW^2)$ as a function of the invariant mass
squared $W^2$ at $Q^2=$10 GeV.
The number of constituents is taken as $n=$2 or 4,
and the cutoff parameter is $\Lambda=\infty$, 1.0, or 0.5 GeV,
and we take $\alpha=\beta=1$.
Here, the hadron $h$ is $f_0$ or $a_0$.
Considering the hadron size and the number of constituents,
we expect that the slope of $W^2$ is much larger if
the hadron has an exotic tetra-quark ($qq\bar q\bar q$)
configuration.}
\label{fig:cross-form}
\end{figure}

Using the form factor together with the GDA expression in
Eq. (\ref{eqn:gda-paramet}), we obtain the cross section
for $e+\gamma \to e + h + \bar h$ as the function of $W^2$
in Fig. \ref{fig:cross-form}.
If the $a_0$ and $f_0$ are ordinary hadrons consist
of $u$ and $d$ quarks
($u\bar d$, $(u\bar u \pm d\bar d)/\sqrt{2}$, $d\bar u$),
the summation over the valence quarks becomes
$\sum_q e_q^2 M_{2\,(q)}^h = (5/18) \sum_a M_{2\,(q)}^h$
by assuming that $M_{2\,(q)}^h$ is flavor independent, whereas 
the factor is $\sum_q e_q^2 M_{2\,(q)}^h = (1/9) \sum_a M_{2\,(q)}^h$
if the $f_0$ is $s\bar s$.
If the $a_0$ and $f_0$ are tetra-quark hadrons
($u\bar s s \bar d$, $(u\bar u \pm d\bar d) s \bar s/\sqrt{2}$, 
$s\bar u d \bar s$),
the factor is 
$\sum_q e_q^2 M_{2\,(q)}^h = (7/18) \sum_a M_{2\,(q)}^h$.
We took into account these factors; however, the $s\bar s$
configuration is not shown in the figure.
The cross sections for $s\bar s$
should be obtained simply by the $n=2$ cross sections 
multiplied by $(2/5)^2$.
The $Q^2$ value should be large enough so that the factorization
into the hard part and the GDAs is satisfied. However, 
if it too large, the cross section becomes too small to
be measured by the current Belle and BaBar experiments. 
As a compromise of these conflicting conditions,
$Q^2$=10 GeV$^2$ is taken in the figure.
The number of constituent for $f_0$ and $a_0$ is taken
as $n=2$ or 4 in the form factor by considering
the ordinary $q\bar q$ or tetra-quark ($qq\bar q\bar q$) state.
The cutoff parameter is taken as $\Lambda=\infty$, 1.0, 
or 0.5 GeV. The cutoff $\Lambda=$1.0 GeV roughly corresponds
to the proton size, and 0.5 GeV is for a more diffuse state.
Just for a typical illustration, we show expected tendency 
by the cutoff and the constituent number $n$
if the hadron $h$ is $q\bar q$ or $qq\bar q\bar q$.
There are five curves and their $W^2$ slopes are distinctly different,
so that the internal configuration of $f_0$ and $a_0$ should be
determined by the cross-section measurements by 
the Belle and BaBar experiments.
If a high-energy linear collider is realized in future,
it has an advantage that larger $W^2$ and $Q^2$ regions
should be probed to clarify the $W^2$ slope.
In any case, the slope data in the region $W^2=$4-6 GeV$^2$ 
regions should be obtained by the current Belle and BaBar.
In this way, the internal structure of exotic hadrons,
for example $f_0$ and $a_0$, should be clarified by
the hadron tomography technique, especially by
using the GDAs.

\vspace{-0.3cm}
\section{Summary}\label{summary}
\vspace{-0.3cm}

It is difficult to find a clear signature of exotic hadrons from
usual low-energy measurements such as masses, spins, parities,
and decay widths. Since the appropriate degrees of freedom
are quarks and gluons at high energies, it is appropriate
to use high-energy hadron reactions, especially by GPDs and GDAs.
In this work, we showed exotic signatures in the GPDs
as the longitudinal momentum distributions and transverse
form factors. We also explained similar exotic signatures
in the GDAs by the functional form of $z$ and 
the energy-momentum form factor. Then, we showed their exotic effects 
on the cross section of $e+\gamma \to e + h + \bar h$.
In particular, we found that there is a distinct signature
of exotic nature in the slope of the invariant-mass squared $W^2$.

The tomography of exotic hadrons is a new promising approach
of exotic hadron studies for clarifying the internal structure 
of exotic hadrons by using the currently developing field of
GPDs and GDAs. 

\vspace{-0.4cm}
\begin{acknowledgements}
\vspace{-0.3cm}

The authors thank 
M. Diehl, V. Guzey, and A. Radyushkin
for communications on GPD parametrization and
S. Uehara for discussions on exotic-hadron studies at Belle.
They also thank W.-C. Chang, J.-C. Peng, and S. Sawada
for communications on J-PARC measurements.
This work was supported by 
the MEXT KAKENHI Grant Numbers 21105006 and 25105010.
\end{acknowledgements}

\vspace{-0.4cm}



\begin{thebibliography}{00}
\bibitem{exotics} R. L. Jaffe, Phys. Rept. {\bf 409}, 1 (2005).
\bibitem{belle} For example, see
          J. Brodzicka {\it et al.}, 
          Prog. Theor. Exp. Phys. {\bf 2012}, 04D001 (2012).
\bibitem{pdg} J. Beringer {\it et al.} (Particle Data Group), 
                      Phys. Rev. D {\bf 86}, 010001 (2012).
\bibitem{spectroscopy-summary} 
     F. E. Close and N. A. T\"ornqvist, J. Phys. G {\bf 28}, R249 (2002);
     C. Amsler and N. A. T\"ornqvist, Phys. Rept. {\bf 389}, 61 (2004).
\bibitem{f0-exotic}
   R. L. Jaffe, Phys. Rev. D {\bf 15}, 267 \& 281 (1977);
   J. D. Weinstein and N. Isgur,
      Phys.\ Rev.\ Lett.\  {\bf 48} (1982) 659;
   D. Black, A. H. Fariborz, and J. Schechter,
      Phys. Rev. D {\bf 61}, 074001 (2000);
   G. 't Hooft, G. Isidori, L. Maiani, A. D. Polosa, V. Riquer,
      Phys. Lett. B {\bf 662}, 424 (2008).
\bibitem{f0-a0-sk} S. Kumano and V. R. Pandharipande, 
                        Phys. Rev. D {\bf 38}, 146 (1988);
                   F. E. Close, N. Isgur, and S. Kumano, 
                        Nucl. Phys. B {\bf 389}, 513 (1993).
\bibitem{sk-2013} T. Sekihara and S. Kumano, arXiv:1311.4637
                        and references therein.
\bibitem{hkos08} M. Hirai, S. Kumano, M. Oka, and K. Sudoh,
                     Phys. Rev. D {\bf 77} (2008), 017504. 
\bibitem{seidl} R. Seidl, personal communications on KEK-B experiments (2012).
\bibitem{kks-2013} H. Kawamura, S. Kumano, and T. Sekihara, 
                       Phys. Rev. D {\bf 88}, 034010 (2013).
\bibitem{other-high-energy}
   M. Diehl, B. Pire, and L. Szymanowski, Phys. Lett. B {\bf 584}, 58 (2004);
   I. V. Anikin, B. Pire, L. Szymanowski, O. V. Teryaev, and S. Wallon,
     Phys. Rev. D {\bf 70}, 011501(R) (2004); {\bf 71}, 034021 (2005)
\bibitem{gpd-summary}
                  K. Goeke, M. V. Polyakov, and M. Vanderhaeghen, 
                  Prog. Part. Nucl. Phys.  {\bf 47}, 401 (2001);
                  X.-D. Ji, Ann. Rev. Nucl. Part. Sci. {\bf 54}, 413 (2004);
                  A. V. Belitsky and A. V. Radyushkin, 
                  Phys. Rept. {\bf 418}, 1 (2005). 
\bibitem{gpd-gda-summary} M. Diehl, Phys. Rept. {\bf 388}, 41 (2003);
         M. Diehl and P. Kroll, Eur. Phys. J. C {\bf 73} (2013) 2397.
\bibitem{transition-gpd} 
   L. L. Frankfurt, M. V. Polyakov, M. Strikman, and M. Vanderhaeghen, 
      Phys. Rev. Lett. {\bf 84}, 2589 (2000).
\bibitem{pire-gpd} E. R. Berger, M. Diehl, and B. Pire,
                          Phys. Lett.  B {\bf 523}, 265 (2001);
   B. Pire and L. Szymanowski, Phys. Lett. B {\bf 622}, 83 (2005);
   J. P. Lansberg, B. Pire, and L. Szymanowski,
         Phys. Rev. D  {\bf 76}, 111502(R) (2007).
\bibitem{kss-gpd} S. Kumano, M. Strikman, and K. Sudoh,
                          Phys. Rev. D {\bf 80}, 074003 (2009).
\bibitem{two-photon} H. Terazawa, Rev. Mod. Phys. {\bf 45}, 615 (1973);
       S. Cooper, Ann. Rev. Nuc. Part. Sci. {\bf 38}, 705 (1988);
       S. Uehara, Nucl. Phys. B (Proc. Suppl.) {\bf 225-227}, 126 (2012).
\bibitem{freund-nlo-dvcs-2001} A. Freund and M. McDermott,
      Eur. Phys. J. C {\bf 23}, 651 (2002).
\bibitem{p-notation} $P=(p+p')/2$ is often used in the GPDs and 
      the same notation $P$ is used in the GDA with the definition
      $P=p+p'$. In order to distinguish them, $\bar P=(p+p')/2$ is used
      in this article.
\bibitem{muller-1994}
   D. M\"uller, D. Robaschik, B. Geyer, F.-M. Dittes, and J. Horejsi, 
       Fortschr. Phys. {\bf 42}, 101 (1994) (hep-ph/9812448).
\bibitem{diehl-2000} M. Diehl, T. Gousset, and B. Pire, 
                        Phys. Rev. Lett. {\bf 81}, 1782 (1998);
                        Phys. Rev. D {\bf 62}, 073014 (2000).
\bibitem{ee-eerhorho}
   I. V. Anikin, B. Pire, and O. V. Teryaev,
     Phys. Rev. D {\bf 69}, 014018 (2004);
     Phys. Lett. B {\bf 626},  86 (2005).
\bibitem{factorization}
   A. Freund, Phys. Rev. D {\bf 61}, 074010 (2000).
\bibitem{personal}
   M. Diehl, V. Guzey, and A. Radyushkin, personal communications (2012).
\bibitem{uehara}
S. Uehara, personal communications (2013).
\bibitem{trans-long} R. Devenish and A. Cooper-Sarkar,
                        Appendix C in {\it Deep Inelastic Scattering}
                        (Oxford University press, 2004).
\bibitem{gpd-paramet}
      M. Vanderhaeghen, P. A. M. Guichon, and M. Guidal, 
            Phys. Rev. D  {\bf 60}, 094017 (1999),
      M. Guidal, M. V. Polyakov, A. V. Radyushkin, and M. Vanderhaeghen,
            Phys. Rev. D {\bf 72}, 054013 (2005);
      V. Guzey, C. Weiss, talks at the GPD working group mini-workshop,
                      JLab, Aug. 6-7, 2008.
\bibitem{radyushkin-paramet}
   A. V. Radyushkin, arXiv:hep-ph/0101225, 
     {\it At the frontier of particle physics : handbook of QCD, }
     edited by M. Shifman (World Scientific, 2001), Vol.2, pp.1038-1099.
   See also Ref. \cite{gpd-gda-summary}.
\bibitem{d-y-west-1970} S. D. Drell and T.-M.Yan, 
                          Phys. Rev. Lett. {\bf 24}, 181  (1970);
                    G. B. West, Phys. Rev. Lett. {\bf 24}, 1206 (1970).
\bibitem{melnitchouk-2005} W. Melnitchouk,  R. Ent, and C. E. Keppel,
                                Phys. Rept. {\bf 406}, 127 (2005).
\bibitem{pqcd-counting} G. R. Farrar and D. R. Jackson, 
                                Phys. Rev. Lett. {\bf 43}, 246 (1979);
         A. V. Efremov and A.V. Radyushkin,
                   Theor. Math. Phys. {\bf 42}, 97 (1980); 
         A. Duncan and A. H. Mueller, Phys. Rev. D {\bf 21}, 1636 (1980); 
                   Phys. Lett. {\bf 93B}, 119 (1980);
      G. P. Lepage and S. J. Brodsky, Phys. Rev. D {\bf 22}, 2157 (1980).
\bibitem{pQCD-large-x-pdf}
         G. F. Farrar and D. R. Jackson, 
                      Phys. Rev. Lett. {\bf 35}, 1416 (1975);
         S. J. Brodsky, M. Burkardt, and I. Schmidt, 
                      Nucl. Phys. B {\bf 441}, 197 (1995).
\bibitem{Q2-code} M. Miyama and S. Kumano, 
                    Comput. Phys. Commun. {\bf 94}, 185 (1996);
	M. Hirai {\it et al.}, 
	                Comput. Phys. Commun. {\bf 108} (1998) 38;
	                {\bf 111}, 150 (1998); 
	                {\bf 183}, 1002 (2012).                 
\bibitem{pion-pdf}
	M. Aicher, A. Sch\"afer, and W. Vogelsang,
                  Phys. Rev. Lett. {\bf 105} (2010) 252003.
    The PDF code was supplied by W. Vogelsang.
\bibitem{proton-pdf}
A. D. Martin, R. G. Roberts, and W. J. Stirling,  
            Phys. Lett. B {\bf 636}, 259 (2006).
\bibitem{form-factors} B. Povh, K. Rith, C. Scholz, and F. Zetsche,
       {\it Particle and Nuclei} (Springer, 1999);
       D. C. Cheng and G. K. O'Neil, 
       {\it Elementary Particle Physics} (Addison-Wesley, 1979).
\bibitem{pdg-kek} J. Beringer {\it et al.} (Particle Data Group), 
                      Phys. Rev. D {\bf 86}, 010001 (2012); 
                      online page http://ccwww.kek.jp/pdg/ .
\bibitem{j-parc-pro-high} W.-C. Chang, J.-C. Peng, S. Sawada {\it et al.},
      An experimental proposal is under investigation for J-PARC.
\bibitem{impact-parameter}
   B. Pire and L. Szymanowski, Phys. Lett. B {\bf 556}, 129 (2003).
\bibitem{gda-sum-rule}
   M. Diehl, T. Feldmann, P. Kroll, and C. Vogt,
     Phys. Rev. D {\bf 61}, 074029 (2000);
   M. V. Polyakov, Nucl. Phys. B {\bf 555}, 231 (1999);
   M. V. Polyakov and C. Weiss, Phys. Rev. D {\bf 60}, 114017 (1999).                  
\bibitem{energy-momentum}
   J. F. Donoghue and H. Leutwyler, Z. Phys. C {\bf 52}, 343 (1991);
   X.-D. Ji, Phys. Rev. Lett. {\bf 78}, 610 (1997); 
   M. V. Polyakov, Phys. Lett. B {\bf 555} 57 (2003);
   H.-C. Kim, P. Schweitzer, and U. Yakhshiev, 
             Phys. Lett. B {\bf 718}, 625 (2012).
\bibitem{BC-1976} S. J. Brodsky and B. T. Chertok, 
             Phys. Rev. D {\bf 14}, 3003 (1976).
\end{thebibliography}
\end{document}